\documentclass[%
 preprint,
 amsmath,amssymb,
 aps,
]{revtex4-2}

\usepackage{graphicx}
\usepackage{dcolumn}
\usepackage{bm}
\usepackage{dsfont}

\usepackage{algorithm}
\usepackage[noend]{algpseudocode}

\usepackage{tikz}
\usetikzlibrary{quantikz}

\begin{document}


\title{Accelerated quantum search using partial oracles and Grover's algorithm}

\author{Fintan Bolton}
\email{fintan.bolton@bradan-quantum.com}
\affiliation{Bradan Quantum, Klugstr. 101, 80637 Munich, Germany}
\affiliation{School of Physics, Trinity College Dublin, Dublin 2, Ireland}



\date{\today}

\begin{abstract}


Grover's algorithm, orginally conceived as a means of searching an unordered database, can also be used to extract solutions from the result sets generated by quantum computations.
The Grover algorithm exploits the concept of an oracle function, which abstracts the process of matching a search item (returning 1 for a match and 0 otherwise), where searching for 1 target item in a search space of size $N$ scales as $\mathcal{O}(\sqrt{N})$ oracle queries.
In this article, we explore the idea of associating a separate oracle with each bit of the matching condition, obtaining multiple \textit{partial oracle functions} which can be tested independently.
Exploring this idea leads to a multi-stage hybrid search algorithm, whose performance can fall within a wide range, anywhere between $\mathcal{O}(\sqrt{N})$ (same as Grover) or $\mathcal{O}(\log(N))$.
The algorithm is validated against the simplest kind of search scenario, where the incoming index bits are scrambled using a permutation operation.

\end{abstract}

\maketitle


\section{Introduction}

Grover's algorithm is an important building block in many quantum algorithms, as it is often needed for searching the result set of a quantum computation.
The $\mathcal{O}(\sqrt{N})$ scaling behaviour discovered by Grover \cite{LKGrover1996,PhysRevLett.79.325} (that is, the number of oracle queries required to find a single target item in a search space of size $N$) thus represents the performance limit for many quantum algorithms.

In the years since its discovery, Grover's algorithm has been extended in various ways. It was soon generalized to cover the case where the target of the search consists of multiple items \cite{AHARONOV_1999,Boyer_Brassard_Hoyer_Tapp_1999}. Given a target set of size $M$, the number of oracle queries scales as $\mathcal{O}(\sqrt{N/M})$.
Moreover, in its original form, Grover's algorithm was not guaranteed to return the target value with 100\% certainty, as the rotations that constitute the algorithm are liable to undershoot or overshoot the target vector. This problem was recognized early on and characterized by Brassard as ``overcooking the souffle" \citep{Brassard1997SearchingAQ}. To solve this problem, work was begun to develop a so-called \textit{fixed-point algorithm} \cite{PhysRevLett.95.150501,grover2006quantum}, which returns a result with 100\% certainty.

Not long afterwards, work by Hoyer \cite{Hoyer_2000} and Long \cite{Long_Li_Zhang_Niu_1999,Long2001GroverAW} led to the discovery of the Grover-Long algorithm, which provides a solution to the fixed-point Grover problem by introducing a single complex phase factor. Later on, multiphase algorithms were also developed \cite{Li2018ComplementarymultiphaseQS,Yoder2014FixedpointQS}, which have the advantage of delivering an accurate result, even when the amplitude of the target state is not known precisely.

In spite of this progress, there are still compelling reasons to look for search algorithms that could outperform Grover's, as highlighted recently in a provocatively titled paper \cite{Stoudenmire2023GroversAO}. The fact that there exists a proof that Grover's search algorithm is optimal \cite{Galindo2000FamilyOG} appears to limit our options.
But there is still room to develop search algorithms that work with different assumptions---in particular, by exploiting specific features of a problem.

For example, in the traditional approach to Grover search, the operation to identify the target of a search is abstracted into an \textit{oracle function}, which returns 1 for a match and 0 otherwise. The oracle function is treated as a black box, returning just 1 bit of information (for a match or a non-match).

However, when we look at specific search problems, there is usually much more than one bit of information available that relates to the matching of target states. For example, consider the algorithm for finding the factors of an integer by Euclidean division. When expressed in binary form, the remainder has multiple bits and every bit of this remainder must be zero for a divisor to be identified as a factor. Do the additional bits of the remainder really have no relevance to the search procedure? This motivates us to consider, more generally, whether multiple bits of information from a match condition can be exploited to improve the search algorithm.

To motivate the discussion, consider a large database indexed by an $x$ field, which contains records with a \texttt{name} field. We want to search the database index $x$ to find the record with \texttt{name} equal to \texttt{Jane Doe}. For the sake of simplicity, we assume that the \texttt{name} field is encoded using a modified Base64 character set, which includes the numerical digits $0$ to $9$, all English uppercase and lowercase letters, the space character (replacing \texttt{+}) and the comma character (replacing \texttt{/}). We assume that all of these characters occur with the same frequency in the \texttt{name} field of the database, so that the letter \texttt{a} occurs with frequency $1/64$, \texttt{b} occurs with frequency $1/64$, and so on.

For this search, the size of the final target set is $1$ (the record matching \texttt{Jane Doe}) and the total number of records in the database is $N=64^m$, where $m$ is the number of Base64 characters in the \texttt{name} field.

Using the conventional Grover algorithm, we would work with the black-box oracle $f(s)$ that returns $1$, if $s$ equals \texttt{Jane Doe}, and $0$, otherwise. In this case, it is known that the Grover algorithm requires $\sqrt{N}=8^m$ oracle queries to complete the search.

Now consider how we could solve the search problem if we had access to the following series of partial oracle functions:
\begin{itemize}
\item $f_J(s)$ returns $1$, if the first letter of $s$ equals \texttt{J}, and $0$ otherwise,
\item $f_{Ja}(s)$ returns $1$, if the first two letters of $s$ equal \texttt{Ja}, and $0$ otherwise,
\item $f_{Jan}(s)$ returns $1$, if the first three letters of $s$ equal \texttt{Jan}, and $0$ otherwise, and so on.
\end{itemize}

We can start by running the Grover algorithm with the partial oracle $f_J(s)$ to search for all names beginning with \texttt{J}. In this case, it is known that the Grover algorithm requires $\sqrt{N/M}$ oracle queries to complete the search, where $M$ is the size of the target set (all $m$-character names beginning with the letter \texttt{J}). Given a \texttt{name} with $m$ Base64 characters, we therefore have $N=64^m$ and $M=64^{m-1}$ (assuming the letter \texttt{J} has the character frequency $1/64$), giving $\sqrt{N/M}=8$ for the number of oracle queries required to complete the first stage of the search.

In the second stage of the search, our starting point is the set of all $m$-character names beginning with the letter \texttt{J}, which gives a starting set of size $N=64^{m-1}$. We work with the partial oracle $f_{Ja}(s)$, which effectively defines a target set of size $M=64^{m-2}$. This gives us a count of $\sqrt{N/M}=8$ for the number of oracle queries required to complete the second stage of the search.

Continuing in this way for all $m$ letters of the \texttt{name} field, we would execute a total of $m$ search stages using the partial oracle approach, for a total oracle query count of $8m$. The difficulty with this approach, however, is that at each stage we are required to restart the search with a different set of index values: first, the set of indices associated with records whose name starts with \texttt{J}; next, the set of indices associated with records whose name starts with \texttt{Ja}; and so on.

In general, enumerating the index values at each stage of the search is a hard problem. In order for this algorithm to be effective, there needs to be an economical way of summarizing the intermediate search results at each stage. This aspect is discussed in section \ref{sec:modelling}.

\section{Algorithm with partial oracles}

Consider the search space addressed by the index $x\in\{0,1\}^n$, which is an $n$-bit binary integer with $2^n=N$ possible values. Define $S$ to be the set of all index values, $S=\lbrace 0,\ldots,N-1\rbrace$.

The search problem we are trying to solve is implicitly defined by a collection of \textit{partial oracles}, which are predicate functions that return either 1 (true) or 0 (false) for each index value $x$. The partial oracles are represented by the partial oracle functions $f_j(x):\{0,1\}^n\rightarrow\lbrace 0,1\rbrace$ for $j=1,\ldots,m$. Associated with each partial oracle function is the partial constraint set $C_j\subset S$, which defines the set of indices for which $f_j(x)$ returns 1:

\begin{equation}
C_j = \{x\,|\,f_j(x)=1\}
\end{equation}

We define the full oracle function $f(x)$ to be the logical conjunction of the partial oracle functions:

\begin{equation}
f(x) = f_1(x)\wedge f_2(x)\wedge \ldots\wedge f_m(x)
\end{equation}

The full oracle function thus returns a value of $1$ only when all of the partial oracle functions are true. We can define the target set $T$ corresponding to the full oracle, which is given by the intersection of all the partial constraints:

\begin{equation}
T = \bigcap_{j=1}^{m}\,C_j
\end{equation}

We can also define the quantum state $|y\rangle$ constructed by mapping the value returned by each partial oracle function to a qubit with 0 or 1 value:

\begin{equation}
|y\rangle = |y_m=f_m(x)\rangle\,\ldots|y_1=f_1(x)\rangle
\end{equation}

We call this $m$-qubit state the \textit{flag state} and the qubit values $|y_j\rangle$ the \textit{flag bits}.

The partial-oracle search algorithm iterates over multiple stages, as the constraints from each partial oracle are successively imposed on the result set: using $f_1(x)$ at stage 1, then $f_1(x)\wedge f_2(x)$ at stage 2, and so on up to $\bigwedge_{j=1}^m f_j(x)$. At each stage, Grover's algorithm is invoked to return a new (smaller) target set.
We can thus define successive target states: $T_1=C_1$ for stage 1, $T_2=C_1\cap C_2$ for stage 2, and for stage $\ell$:

\begin{equation} \label{eq:targetsetdefn}
T_\ell = \bigcap_{j=1}^{\ell}\,C_j
\end{equation}

At the last stage $m$, we get the final target state $T_m=T$, and for notational convenience we also set $T_0 = S$, which represents the initial state of the system. It is also useful to define the corresponding target set sizes, $M_\ell = |T_\ell|$, and for notational convenience, we also define $M_0=|S|=N$. As the search algorithm progresses, the target sets get successively smaller, so we find that:

\begin{equation}
M_0 > M_1 > M_2 > \ldots > M_m
\end{equation}

For example, if we are searching for a unique item in the search space, we would start with $M_0=N$ and finish with $M_m=1$.

Before presenting the details of the iterative algorithm, we find it useful to define the following approach for mapping an arbitrary index set $X\subset S$ to a quantum state $|X\rangle$:

\begin{equation} \label{eq:indexstate}
X\,\,\longmapsto\,\,  |X\rangle = \frac{1}{\sqrt{|X|}}\,\, \sum_{x^\prime\in X}\,|x^\prime\rangle
\end{equation}

Where $|x^\prime\rangle$ is the of qubit state corresponding to the $x^\prime$ index. Note that every index state in this sum acquires the $+1$ phase, so that the Hilbert vector $|X\rangle$ points diagonally with a component in the positive direction of every $|x\rangle$ axis.

We now present an outline of the iterative algorithm based on partial oracles, which invokes the Grover algorithm at each stage. In the first stage, the initial index state $|S\rangle$ is an equal superposition state, prepared by applying the Walsh-Hadamard transformation to all of the zeroed index qubits:

\begin{equation}
|S\rangle = H^{\otimes n}\, |0\rangle = \frac{1}{\sqrt{N}}\,\,\sum_{i=1}^{N} |x_i\rangle
\end{equation}

In the first stage, we work with the partial oracle $f_1(x)$, which implies we are searching for the target set $T_1=C_1$ of size $M_1$. The initial state $|S\rangle$ can be written as the sum of the two orthogonal vectors, $|S\setminus C_1\rangle$ and $|C_1\rangle$ (defined using equation \ref{eq:indexstate}), as follows:

\begin{equation}
|S\rangle = \sqrt{1-M_1/N}\,\,|S\setminus C_1\rangle + \sqrt{M_1/N}\,\,|C_1\rangle
\end{equation}

Following the notational conventions in \cite{Yoder2014FixedpointQS}, we introduce $\lambda_1=M_1/N$ and $\overline{\lambda}_1=1-M_1/N$, so that we can rewrite this equation as:

\begin{equation}
|S\rangle = \sqrt{\overline{\lambda}_1}\,\,|S\setminus C_1\rangle + \sqrt{\lambda_1}\,\,|C_1\rangle
\end{equation}

This equation represents the usual starting point for invoking Grover's algorithm. By invoking Grover's algorithm with $f_1(x)$ as the oracle, the initial state $|S\rangle$ is rotated until it aligns with the target state $|T_1\rangle$ (equal to $|C_1\rangle$ by equation \ref{eq:targetsetdefn}). After invoking Grover's algorithm, the new state of the system is $|T_1\rangle$ and the amplitude of the orthogonal state is zero (ignoring errors, and assuming we use a deterministic version of Grover's algorithm).

In the second stage, we begin with $|T_1\rangle$ as the initial state and work with the oracle $f_1(x)\wedge f_2(x)$, which implies we are searching for the target set $T_2=T_1\cap C_2$ of size $M_2$. The initial state $|T_1\rangle$ can be written as the sum of the two orthogonal vectors, $|T_1\setminus C_2\rangle$ and $|T_1\cap C_2\rangle$, as follows:

\begin{equation}
|T_1\rangle = \sqrt{\overline{\lambda}_2}\,\,|T_1\setminus C_2\rangle + \sqrt{\lambda_2}\,\,|T_1\cap C_2\rangle
\end{equation}

Where we have defined $\lambda_2=M_2/M_1$ and $\overline{\lambda}_2=1-M_2/M_1$. The outcome of invoking Grover's algorithm with oracle $f_1(x)\wedge f_2(x)$ and initial state $|T_1\rangle$ is to rotate the state of the system to the target state $|T_2\rangle=|T_1\cap C_2\rangle$.

Generalizing this to the $\ell^\textrm{th}$ step, we can write:

\begin{equation} \label{eq:lthstep}
|T_{\ell-1}\rangle = \sqrt{\overline{\lambda}_\ell}\,\,|T_{\ell-1}\setminus C_\ell\rangle + \sqrt{\lambda_\ell}\,\,|T_{\ell-1}\cap C_\ell\rangle
\end{equation}

Where we have defined $\lambda_\ell=M_\ell/M_{\ell-1}$ and $\overline{\lambda}_\ell=1-M_\ell/M_{\ell-1}$. The outcome of invoking Grover's algorithm with oracle $\bigwedge_{j=1}^{\ell}f_j(x)$ and initial state $|T_{\ell-1}\rangle$ is to rotate the state of the system to the target state $|T_\ell\rangle=|T_{\ell-1}\cap C_\ell\rangle$, which consists of index states that satisfy the partial constraints $f_1(x)\wedge\ldots\wedge f_\ell(x)=1$.

Note that the idea that Grover's algorithm can work with an initial state $T_{\ell-1}$ that is not an equal superposition state (given by $H^{\otimes n}\ket{0}$) is not new. Other authors have used this idea to prepare an optimized initial state \cite{Tulsi2016OnTC,Singleton2021GroversAW} and demonstrated that the approach is compatible with Grover's algorithm. 

Now we turn our attention to the particular form of Grover's algorithm that is invoked at each stage. In principle, it is possible to use any variant or generalization of Grover's algorithm. But the Grover-Long algorithm \cite{Long2001GroverAW} is a good choice in this context because it is a fixed-point algorithm and relatively simple to implement.

The notation we use to describe the Grover-Long algorithm roughly follows \cite{Yoder2014FixedpointQS}. But when it comes to defining the reflection operator, our notation uses a formal template, because we require multiple operators of this type. Given the set of index states $X$ and the angle $\alpha$, we define the template for the reflection operator $\mathcal{S}(\alpha, X)$, as follows:

\begin{equation}
\mathcal{S}(\alpha, X) = \mathbf{I} + (e^{i\alpha}-1)\,|X\rangle \langle X|
\end{equation}

Where the $|X\rangle$ state is defined using equation \ref{eq:indexstate}.

Consider how to invoke the Grover-Long algorithm to solve the $\ell^{th}$ iteration of the algorithm (see equation \ref{eq:lthstep}). Define the following orthogonal basis states:

\begin{equation}
|T_{\ell-1}\setminus C_\ell\rangle \longmapsto \left(
\begin{array}{l}
1 \\
0
\end{array}
\right)
\quad \textrm{and} \quad
|T_{\ell-1}\cap C_\ell\rangle \longmapsto \left(
\begin{array}{l}
0 \\
1
\end{array}
\right)
\end{equation}

Write the initial state $|T_{\ell-1}\rangle$ in terms of this basis:

\begin{equation}
|T_{\ell-1}\rangle = 
\left(
\begin{array}{l}
\sqrt{\overline{\lambda}_\ell} \\
\sqrt{\lambda_\ell}
\end{array}
\right)
\end{equation}

The first reflection operator we need for the Grover-Long algorithm is used to mark the target state $|T_\ell\rangle$ (equal to $|T_{\ell-1}\cap C_\ell\rangle$) with the complex phase $e^{i\alpha}$:

\begin{eqnarray*}
\mathcal{S}(\alpha, T_\ell) &=& \mathbf{I} + (e^{i\alpha}-1)\,|T_\ell\rangle \langle T_\ell| \\
&=&
\left(
\begin{array}{ll}
1 & 0 \\
0 & e^{i\alpha}
\end{array}
\right)
\end{eqnarray*}

The second reflection operator we need for the Grover-Long algorithm is a reflection about the initial state $|T_{\ell-1}\rangle$ marked with the complex phase $e^{i\alpha}$:

\begin{eqnarray} \label{eq:diffusionop}
\mathcal{S}(\alpha, T_{\ell-1}) &=& \mathbf{I} + (e^{i\alpha}-1)\,|T_{\ell-1}\rangle \langle T_{\ell-1}| \\
&=&
\left(
\begin{array}{ll}
1+(e^{i\alpha}-1)\overline{\lambda_\ell} & \quad (e^{i\alpha}-1)\sqrt{\lambda_\ell\overline{\lambda}_\ell} \\
(e^{i\alpha}-1)\sqrt{\lambda_\ell\overline{\lambda}_\ell} & \quad 1+(e^{i\alpha}-1)\lambda_\ell
\end{array}
\right)
\end{eqnarray}

The product of these two reflection operators gives the Grover iterate for the $\ell^\textrm{th}$ stage of the outer algorithm loop:

\begin{equation}
G_\ell(\alpha) = - \mathcal{S}(\alpha, T_{\ell-1})\,\mathcal{S}(\alpha, T_\ell)
\end{equation}

Now we need to figure out how many times to apply the Grover operator $G_\ell(\alpha)$, in order to reach the target state. According to Grover-Long \cite{Long2001GroverAW}, the number of required iterations $g_\ell$ depends on $\lambda_\ell$ and is given by:

\begin{equation}\label{eq:gl}
g_\ell = \left\lceil \frac{\pi}{4\arcsin\sqrt{\lambda_\ell}} - \frac{1}{2} \right\rceil
\end{equation}

The angle that defines the complex phases is given by:

\begin{equation}\label{eq:alpha}
\alpha_\ell = 2\arcsin\left( \frac{1}{\sqrt{\lambda_\ell}} \sin\left( \frac{\pi}{4 g_\ell + 2} \right) \right)
\end{equation}

Combining the iterations from the inner loop (Grover-Long algorithm) and the outer loop (partial oracle stages), the complete algorithm can be summarized in terms of the Grover operators, as follows:

\begin{equation}
\left(G_m(\alpha_m)\right)^{g_m}\ldots\left(G_\ell(\alpha_\ell)\right)^{g_\ell}\ldots\left(G_1(\alpha_1)\right)^{g_1}\>\ket{T_0} = \ket{T_m}
\end{equation}

\section{Algorithm performance}\label{sec:algoperformance}

\subsection{Theoretical optimum}

The partial oracle algorithm is organised in a sequence of $m$ stages, where the size of the target set $M_\ell$ becomes successively smaller, $M_0 > \ldots > M_\ell > \ldots > M_m$.
At each stage, $\ell$, the partial oracle algorithm invokes the Grover-Long algorithm with $g_\ell$ oracle queries. Using equation \ref{eq:gl}, we can write the total number of oracle queries as:

\begin{equation}
\sum_{\ell=1}^m g_\ell =
\sum_{\ell=1}^m \left\lceil \frac{\pi}{4\arcsin\sqrt{M_\ell/M_{\ell-1}}} - \frac{1}{2} \right\rceil
\end{equation}

From this equation, we see that the theoretically optimum solution of the partial oracle algorithm depends \textit{only} on the sequence of target set sizes $\lbrace M_1,\ldots, M_m\rbrace$.

The individual terms $g_\ell$ of this sum tend to become small when $M_\ell$ and $M_{\ell-1}$ are roughly comparable in size. In particular, for $1/4 < M_\ell/M_{\ell-1} < 1/2$ we get $g_\ell=1$ and, if this holds for all $\ell$, the total number of oracle queries is just $m$. The fastest reduction in the size of the target set size is thus attained when $M_\ell/M_{\ell-1}=1/4$ for all $\ell$.

We can now compare the partial oracle algorithm with the conventional Grover algorithm for this theoretically optimum case, where $M_\ell/M_{\ell-1}=1/4$ for all $\ell$. For the partial oracle algorithm, the number of required oracle queries is just $m$. For the conventional Grover algorithm (where $N=M_0$ is the number of initial states and $M=M_m$ is the number of target states) the number of required oracle queries is $\sqrt{N/M}=\sqrt{\prod_{\ell=1}^m (M_{\ell-1}/M_\ell)}=2^m$. 

In practice, achieving this theoretical optimum depends heavily on the efficient modelling of intermediate states $|T_\ell\rangle$ (see section \ref{sec:modelling}). If an intermediate state cannot be modelled efficiently, this increases $g_\ell$ and, in the worst case, the algorithm reverts to the same number of oracle queries as the conventional Grover algorithm.

\subsection{Case of one-to-one mapping between flag and index}\label{sec:onetoone}

To illustrate the relevance of the preceding performance analysis, we provide an example of a class of search problems that satisfy the prerequisites for a theoretically optimum search.

Consider a search problem characterised by the fact that there is a \textit{one-to-one mapping} between the index states and the partial oracle flag states. That is, there exists a bijective function $F:\lbrace 0,1\rbrace^n\mapsto\lbrace 0,1\rbrace^n$ that maps $n$-bit index states $x$ to $n$-bit flag states $y$. Applying the partial oracle algorithm to this problem gives a problem with $m=n$ stages, assuming we constrain the search with one additional flag bit at each stage.

We can calculate the target set sizes $\lbrace M_\ell\rbrace_{\ell=1}^n$ as follows. The number of initial states $M_0=2^n$. Let us write the flag state in bitwise form as $y=y_n\ldots y_1$. When the first flag bit $y_1=1$, the flag values are constrained to be $y=y_n\ldots y_2 1$, which can take a total of $2^{n-1}$ different values. Because the function $F(x)$ is bijective, this corresponds to a set of $2^{n-1}$ distinct $x$ values that satisfy the constraint $y_1=1$, which implies that $M_1=2^{n-1}$.
Similarly, with the first two flag bits set to $y_2=y_1=1$, the flag values are constrained to be $y=y_n\ldots y_3 11$, corresponding to a target set size of $M_2=2^{n-2}$. Continuing in this way, we find that $M_\ell=2^{n-\ell}$ for each stage $\ell$ of the search problem. Thus we have $M_{\ell}/M_{\ell-1}=1/2$ for all values of $\ell$, which implies that $g_\ell=1$ at each stage.

Note that in this case we have a choice between working with one additional flag bit at each stage of the algorithm, giving a solution in $n$ steps, or working with two additional flag bits at each stage, giving a solution in $n/2$ steps.

\subsection{Modelling intermediate states}\label{sec:modelling}

In order to get the best performance out of the partial oracle algorithm, we need to be able to model the intermediate states $\ket{T_\ell}$ efficiently. To see why this is so, consider what we need to do to construct the Grover circuit for stage $\ell$ of the algorithm. The Grover diffusion operator for stage $\ell$ is given by equation \ref{eq:diffusionop}. In order to construct a circuit for the Grover diffusion operator, we need to write it as a product of three operators:

\begin{equation}
\mathcal{S}(\alpha, T_{\ell-1}) = R(T_{\ell-1})\left(\mathbf{I} + (e^{i\alpha}-1)\,\ket{0} \bra{0} \right) R^\dagger(T_{\ell-1})
\end{equation}

Where $\ket{T_{\ell-1}}$ is an intermediate state that is the initial state of stage $\ell$, and $R(T_{\ell-1})$ is a unitary operator that rotates the $\ket{0}$ state to the $\ket{T_{\ell-1}}$ state:

\begin{equation}
R(T_{\ell-1})\,\ket{0} = \ket{T_{\ell-1}}
\end{equation}

The problem is that we need to build an explicit circuit that performs the operation $R(T_{\ell-1})$ (and the adjoint $R^\dagger(T_{\ell-1})$) and this is generally difficult to do for an arbitrary intermediate state $\ket{T_{\ell-1}}$. In the special case where the starting point is an equal superposition of states, the operator reduces to a Walsh-Hadamard transformation, $R(S)=H^{\otimes n}$. But in the general case we need to construct an approximation to $R(T_{\ell-1})$ by basing the operator on a modelled state.

A modelled state $\ket{\mu_\ell}$ is an approximation to a target state, $\ket{\mu_{\ell}}\approx \ket{T_{\ell}}$, that can be written compactly, where the goal is for the modelled state to reproduce (approximately) the same basis state probabilities as the true target state, $\ket{T_\ell}$. We can construct the modelled state $\ket{\mu_\ell}$ by measuring the state of the index qubits after the $\ell^\textrm{th}$ Grover-Long stage, over a set of $N_\textrm{shot}$ shots, and then performing a statistical analysis of the results. We could extract bitwise correlations (the probability of measuring a $1$ instead of a $0$ for a particular bit), and/or two-bit correlations, or higher-order correlations, as necessary.

Given a compact description of the modelled state $\ket{\mu_\ell}$, it becomes possible to construct a circuit for the corresponding rotation operator $R(\mu_\ell)$, which satisfies $R(\mu_\ell)\,\ket{0}=\ket{\mu_\ell}$. This rotation operator can then be used to construct a circuit that approximates the Grover-Long diffusion operator for the next stage.

Note that the potential difficulty of constructing modelled states represents one of the key challenges of the partial oracle algorithm. If state modelling requires you to take into account many high-order bit correlations, it becomes unwieldy. The algorithm works best in cases where relatively low-order correlations are adequate to model the intermediate states.

\section{Algorithm in practice}

\subsection{Example scenario}\label{sec:scrambler}

We now consider a simple search problem, the \textit{scrambler scenario}, with the aim of testing and validating the partial oracle algorithm.
The scrambler scenario is set up in a similar way to an algorithm for cracking an encryption key. Given a matching plaintext $p_0$ and ciphertext $c_0$ pair, the search problem consists of looking for the key $x$ that satisfies:

\begin{equation}
\textrm{Scram}(p_0, x) = c_0
\end{equation}

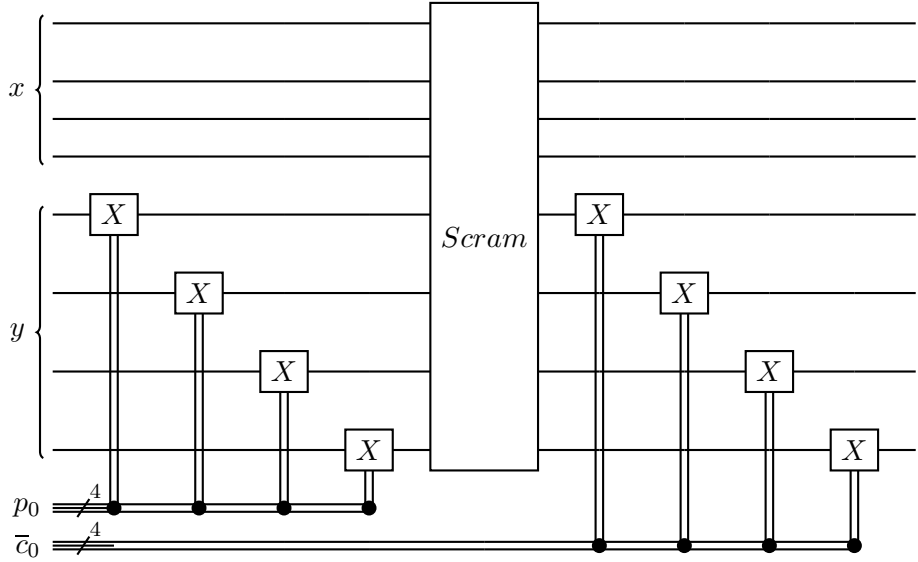
\begin{figure}
\begin{tikzcd}
\lstick[wires=4]{$x$} & \qw & \qw & \qw & \qw & \gate[8]{\text{$Scram$}} & \qw & \qw & \qw & \qw & \qw \\
& \qw & \qw & \qw & \qw & & \qw & \qw & \qw & \qw & \qw \\
& \qw & \qw & \qw & \qw & & \qw & \qw & \qw & \qw & \qw \\
& \qw & \qw & \qw & \qw & & \qw & \qw & \qw & \qw & \qw \\
\lstick[wires=4]{$y$} & \gate{X}\vcw{4} & \qw & \qw & \qw & & \gate{X}\vcw{5} & \qw & \qw & \qw & \qw \\
& \qw & \gate{X}\vcw{3} & \qw & \qw & & \qw & \gate{X}\vcw{4} & \qw & \qw & \qw \\
& \qw & \qw & \gate{X}\vcw{2} & \qw & & \qw & \qw & \gate{X}\vcw{3} & \qw & \qw \\
& \qw & \qw & \qw & \gate{X}\vcw{1} & & \qw & \qw & \qw & \gate{X}\vcw{2} & \qw \\
\lstick{$p_0$} & \qwbundle{4}\cwbend{0} & \cwbend{0} & \cwbend{0} & \cwbend{0} &  &  &  &  &  &  \\
\lstick{$\overline{c}_0$} & \qwbundle{4}\cw & \cw & \cw & \cw & \cw & \cwbend{0} & \cwbend{0} & \cwbend{0} & \cwbend{0} & \\
\end{tikzcd}

\caption{Scrambler circuit over 4-qubit index}\label{fig:scrambler}
\end{figure}

The circuit for the scrambler scenario is shown in figure \ref{fig:scrambler}, where it is assumed that the $x$ qubits have already been initialised by applying the Walsh-Hadamard transformation (so that the $x$ register is in an equal superposition of all possible key values). The given plaintext $p_0$ is written to the $y$ register using CNOT gates. The Scram operator takes the key value $x$ and plaintext $p_0$ as inputs and generates the corresponding ciphertext $c=\textrm{Scram}(p_0,x)$ on the $y$ register. Finally, the given ciphertext $c_0$ is compared (negative XORed) with the generated ciphertext $c$ to create the flag bits on the $y$ register. If there is a perfect match between $c$ and $c_0$, all of the flag bits are $1$.

The Scram operator has the following notable characteristics: it induces a one-to-one mapping between the $n$-bit flag state and the $n$-bit index state, so that the prerequisites for a theoretically optimum search are satisfied (see section \ref{sec:onetoone}).
It uses only bitwise operations to scramble the plaintext (swapping bits around and XORing the plaintext with the key) so that a simple bitwise approach can be used for the modelled states (see section \ref{sec:modelling}).

\subsection{Algorithm outline}

The following algorithm outline summarizes the steps for the partial oracle algorithm:

\begin{algorithm}[H]
\caption{Outline of Grover algorithm with partial oracles}
\begin{algorithmic}

\Procedure{GroverWithPartialOracle}{}
    \For{$\ell\gets 1, m$}    \Comment{Iterate over $m$ stages}
        \State Step 1: Define the partial oracle for stage $\ell$: $\bigwedge_{j=1}^{\ell}\,f_j(x)$
        \State Step 2: Get the initial state: $|T_0\rangle=H^{\otimes n}\,|0\rangle\>$ or $\>|T_{\ell-1}\rangle\approx |\mu_{\ell-1}\rangle$
        \State Step 3: Construct operator $R(\mu_{\ell-1})$, such that $R(\mu_{\ell-1})\,|0\rangle=|\mu_{\ell-1}\rangle$
        \State Step 4: Calculate/estimate weight of the target state: $\lambda_\ell\gets M_\ell/M_{\ell-1}$
        \State Step 5: Define the phase oracle: $U_f \gets \mathbf{I} + (e^{i\alpha}-1)\,|T_\ell\rangle \langle T_\ell|$
        \State Step 6: Run the Grover-Long algorithm with: $\lambda_\ell$, $U_f$, and $R(\mu_{\ell-1})$
        \State Step 7: Generate the next modelled state $|\mu_\ell\rangle$ from the Grover result
    \EndFor
\EndProcedure

\end{algorithmic}
\end{algorithm}

\subsection{Algorithm implementation}

At each stage of the algorithm, we include more of the constraints defined by the partial oracles, adding one partial oracle constraint at each stage. Thus in stage 1 we work with $f_1(x)$, in stage 2 we work with $f_1(x)\wedge f_2(x)$, and in stage $\ell$ we work with $\bigwedge_{j=1}^\ell f_j(x)$. In the case of the scrambler example, each partial oracle $f_j(x)$ flags a single matching bit between the generated scrambled text and the expected scrambled text.

In the first stage of the algorithm, the initial state is got by applying the Walsh-Hadamard transformation to the $|0\rangle$ state, giving the state $|T_0\rangle=H^{\otimes n}\,|0\rangle\>$. In subsequent stages of the algorithm, the initial state must be obtained from a modelled state, $|\mu_{\ell-1}\rangle\approx|T_{\ell-1}\rangle$. Because the scrambler example only introduces bitwise correlations between the flag states $y$ and the index states $x$, it is sufficient to use a bitwise modelled state for this case. The bitwise modelled state $|\mu\rangle$ has the following form:

\begin{equation}\label{eq:modelledstatedefn}
|\mu\rangle\,=\,\left(\cos(\beta_n/2)|0\rangle+\sin(\beta_n/2)|1\rangle\right)\ldots\left(\cos(\beta_1/2)|0\rangle+\sin(\beta_1/2)|1\rangle\right)
\end{equation}

Where each bit can be biased to favour either the $|0\rangle$ state (with 100\% probability, when $\beta_j=0$) or the $|1\rangle$ state (with 100\% probability, when $\beta_j=\pi$), and the parameters $\beta_1,\ldots,\beta_n$ are adusted to give the best fit for the state we are trying to model.

Corresponding to the modelled state $|\mu\rangle$, we need to construct a rotation operator $R(\mu)$ that can rotate from the zero state to the modelled state, as in $R(\mu)\,|0\rangle=|\mu\rangle$. Because we are working with a bitwise modelled state, we can write the rotation operator as a product of bitwise operators, $R(\mu)=R_{\beta_n}R_{\beta_{n-1}}\ldots R_{\beta_1}$, where each $R_{\beta_j}$ is defined as:

\begin{equation}
R_{\beta_j}=
\left(
\begin{array}{rr}
\cos(\beta_j/2)\>\> & \sin(\beta_j/2) \\
\sin(\beta_j/2)\>\> & -\cos(\beta_j/2)
\end{array}
\right)
\end{equation}

In the special case of $\beta_j=\pi/2$, this corresponds to a Hadamard operator, $R_{\pi/2}=H$. Note also that $R_{\beta_j}=R_{\beta_j}^\dagger$ and $[R_{\beta_i},R_{\beta_j}]=0$, so we have $R(\mu)=R^\dagger(\mu)$.

To estimate the weight of the next target state $\lambda_\ell$, we need the Shannon entropy of the $\mu_{\ell-1}$ modelled state. Applying the definition of Shannon entropy to equation \ref{eq:modelledstatedefn}, we get:

\begin{equation}
H(\mu_{\ell-1}) = \sum_{j=1}^m \left(
-\cos^2(\beta_j/2)\log_2(\cos^2(\beta_j/2))
-\sin^2(\beta_j/2)\log_2(\sin^2(\beta_j/2))
\right)
\end{equation}

It can then be shown that, if the modelled state entropy deviates only slightly from the ideal target entropy $H(T_{\ell-1})=n-(\ell-1)$, the weight of the next target state can be approximated by $\lambda_\ell \approx 2^{((n-l)-H(\mu_{\ell-1}))}$, which gives values $\lambda_\ell\approx 0.5$ for well-modelled states.

The phase oracle $U_f$ is the operator that defines the first step of the Grover algorithm. The abstract representation of the phase oracle is given by:

\begin{equation}
U_f \equiv \mathcal{S}(\alpha,T_\ell) = \mathbf{I} + (e^{i\alpha}-1)\,|T_\ell\rangle \langle T_\ell|
\end{equation}

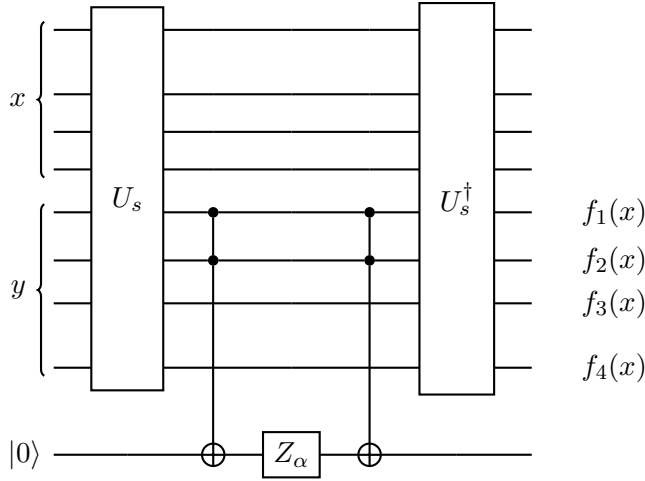
\begin{figure}
\begin{tikzcd}
\lstick[wires=4]{$x$} & \gate[8]{\text{  $U_s$  }} & \qw & \qw & \qw & \gate[8]{\text{  $U_s^\dagger$  }} & \qw \\
& & \qw & \qw & \qw & & \qw \\
& & \qw & \qw & \qw & & \qw \\
& & \qw & \qw & \qw & & \qw \\
\lstick[wires=4]{$y$} & & \ctrl{1} & \qw & \ctrl{1} & & \qw & \rstick{$f_1(x)$} \\
& & \ctrl{3} & \qw & \ctrl{3} & & \qw & \rstick{$f_2(x)$} \\
& & \qw & \qw & \qw & & \qw & \rstick{$f_3(x)$} \\
& & \qw & \qw & \qw & & \qw & \rstick{$f_4(x)$} \\
\lstick{\ket{0}} & \qw & \targ{} & \gate{Z_\alpha} & \targ{} & \qw & \qw \\
\end{tikzcd}

\caption{Phase oracle circuit for $f_1(x)\wedge f_2(x)$}\label{fig:oracle}

\end{figure}

Where $|T_\ell\rangle$ represents the target state for this stage of the algorithm and $\alpha$ is the phase for marking the target states. The circuit for $U_f$ is shown in figure \ref{fig:oracle}.

The phase oracle circuit $U_f$ incorporates the complete scrambler circuit from figure \ref{fig:scrambler}, encapsulating it in the box labelled $U_s$, and adds the gates for marking the target states with the $e^{i\alpha}$ phase (shown with the $f_1(x)\wedge f_2(x)$ oracle function in figure \ref{fig:oracle}). Note the importance of applying the inverse scrambler circuit $U_s^\dagger$ at the end, which disentangles the x qubits from the y qubits.

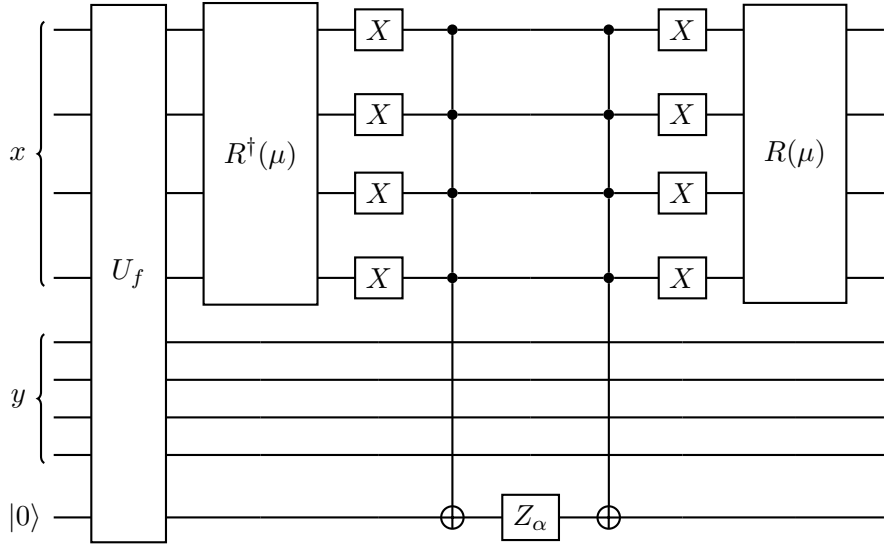
\begin{figure}
\begin{quantikz}
\lstick[wires=4]{$x$} & \gate[9]{\text{  $U_f$  }} &  \gate[4]{\text{  $R^\dagger(\mu)$  }} & \gate{X} & \ctrl{1} & \qw & \ctrl{1} & \gate{X} & \gate[4]{\text{  $R(\mu)$  }} & \qw \\
& & & \gate{X} & \ctrl{1} & \qw & \ctrl{1} & \gate{X} & & \qw \\
& & & \gate{X} & \ctrl{1} & \qw & \ctrl{1} & \gate{X} & & \qw \\
& & & \gate{X} & \ctrl{5} & \qw & \ctrl{5} & \gate{X} & & \qw \\
\lstick[wires=4]{$y$} & & \qw & \qw & \qw & \qw & \qw & \qw & \qw & \qw \\
& & \qw & \qw & \qw & \qw & \qw & \qw & \qw & \qw \\
& & \qw & \qw & \qw & \qw & \qw & \qw & \qw & \qw \\
& & \qw & \qw & \qw & \qw & \qw & \qw & \qw & \qw \\
\lstick{\ket{0}} & & \qw & \qw & \targ{} & \gate{Z_\alpha} & \targ{} & \qw & \qw & \qw \\
\end{quantikz}

\caption{Grover-Long circuit for one iteration of $G_\ell(\alpha_\ell)$}\label{fig:groverlong}

\end{figure}

At this point, we have all the prerequisites for defining the Grover-Long circuit, which depends on $\lambda_\ell$, $U_f$, and $R(\mu_{\ell-1})$. First we invoke equation \ref{eq:gl} to find out how many Grover iterations $g_\ell$ are required, and then equation \ref{eq:alpha} to find the amplitude phase factor $\alpha_\ell$ for this stage. Figure \ref{fig:groverlong} shows a \textit{single} iteration of the Grover-Long algorithm.

The first step of the Grover-Long algorithm is to apply the oracle phase operator $U_f$ (which encapsulates the circuit shown in figure \ref{fig:oracle}). The second step is to apply the Grover diffusion operator, implemented by the rest of the circuit shown in figure \ref{fig:groverlong}.
The diffusion operator starts by applying the $R^\dagger(\mu)$ inverse rotation, followed by a series of gates that apply the $e^{i\alpha}$ phase to the $|0\rangle$ state, and then the diffusion operator ends by applying the $R(\mu)$ rotation.

Running the Grover-Long circuit takes us from the initial state $|T_{\ell-1}\rangle$ (approximated by $|\mu_{\ell-1}\rangle$) to the target state $|T_\ell\rangle$. To create the next modelled state $|\mu_\ell\rangle$, however, we need to run this circuit repeatedly (multiple shots) to get statistics that we can analyse. For a bitwise modelled state, the analysis is simple. The probability that $|x_j\rangle$ equals $|1\rangle$ can be found simply by counting how many times the $|1\rangle$ state occurs and dividing it by the total number of shots. In other words:

\begin{equation}
\textrm{Probability that }|x_j\rangle\textrm{ equals }|1\rangle =
\frac{\textrm{\#Count of }|x_j=1\rangle}{\textrm{\#Shots}}
\end{equation}

From equation \ref{eq:modelledstatedefn}, we see that this probability corresponds to $\sin^2(\beta_j/2)$, which gives us the modelling parameter $\beta_j$. However, we need to be cautious about setting $\beta_j=0$ (100\% probability of $|0\rangle$ state) or $\beta_j=\pi$ (100\% probability of $|1\rangle$ state), because we don't want to completely exclude the possibility of an opposite bit value, given that the modelled state is based on approximate statistics from a limited number of shots, $N_\textrm{shots}$. From the theory of binomial distributions, the recommended minimum probability in this case is given by the Bayes estimator $p_{min}=1/(1+N_\textrm{shots})$.

\subsection{Results}

The partial oracle algorithm was run against the Scrambler example (section \ref{sec:scrambler}), with the aim of validating the algorithm against a simple search scenario (see sample code in \cite{Bolton_Partial_oracle_search_2024}). Because the Scrambler circuit performs only reversible bitwise operations (XORing the key value with the plaintext, swapping bits around), it induces a simple one-to-one mapping between the flag bits and the search index. As explained in section \ref{sec:onetoone}, this implies that the Scrambler search problem satisfies the prerequisites for optimal search performance (comparable to binary search).

Trial runs of the Scrambler search problem confirm that it can be solved with $\mathcal{O}(\log(N))$ oracle queries. For example, for a key length of 14 qubits, the partial oracle algorithm returns a solution after just 14 oracle queries (not counting the repetitions from multiple shots), whereas a plain Grover search would require $2^7=128$ oracle queries.

When measuring the the key value (search index $x$), the probability of obtaining the correct target value was usually $>0.90$, but occasionally significantly lower. Preliminary inspection of the results suggested that this was due to statistical inaccuracies in the modelled state. It should be possible to reduce this inaccuracy by increasing the number of shots $N_\textrm{Shots}$ (which gives better state modelling). Table \ref{tab:results} shows the results of solving the Scrambler search problem with an 8-qubit key for increasing values of $N_\textrm{Shots}$. As can be seen from the table, the target accuracy improves considerably as $N_\textrm{Shots}$ increases.

\begin{table}
\begin{ruledtabular}
\begin{tabular}{rcccccc}
$N_\textrm{Shots}$ &  &  & \textbf{Probability ranges} &  &  & \\
\hline
 & 0.50--0.59 & 0.60--0.69 & 0.70--0.79 & 0.80--0.89 & 0.90--0.99 & $>0.99$ \\
\hline
 200 & 1 & 0 & 0 & 1 & 9 & 9 \\
 400 & 0 & 0 & 0 & 0 & 6 & 14 \\
 600 & 0 & 0 & 0 & 0 & 3 & 17 \\
 800 & 0 & 0 & 0 & 0 & 1 & 19 \\
1000 & 0 & 0 & 0 & 0 & 2 & 18
\end{tabular}
\end{ruledtabular}
\caption{Probability of returning the correct search result from the Scrambler search scenario with an 8-qubit key length. For each value of $N_\textrm{Shots}$, the partial oracle algorithm was repeated 20 times and the columns show how often a particular success probability was achieved.}
\label{tab:results}
\end{table}

\section{Conclusion}


There are many search problems in the real world that provide multiple bits of information related to matching the target of a search. In the traditional approach to quantum search using Grover's algorithm, you would wrap these bits of information with a single oracle function that provides just one bit of output. In this article, however, we have shown that an alternative approach is possible, where the additional bits of information are represented as partial oracle functions. The resulting partial oracle search is a multi-stage algorithm, where additional constraints from the partial oracles are added at each stage. The idea that multiple oracles can be used to accelerate quantum search is also supported by recent work by Dash et all \cite{Dash2023BiQuadraticII} and Khadiev and Krendeleva \cite{Khadiev2023QuantumAF}, where two oracles from different systems are combined to give a search algorithm that outperforms a simple application of Grover's algorithm.

We have shown that the performance of the partial oracle algorithm depends on the following aspects of a given search problem:

\begin{itemize}
\item The way in which the target set size $M_\ell$ decreases as the partial oracle constraints are successively applied (see equation \ref{eq:gl}). In particular, optimum search performance is possible when $1/4 < M_\ell/M_{\ell-1} < 1/2$ holds for all $\lbrace M_\ell\rbrace$

\item The feasibility of representing intermediate search results by a modelled state that approximates the ideal target state $|\mu_{\ell-1}\rangle\approx|T_{\ell-1}\rangle$. If the modelled state has a low accuracy, this affects the weight of the target state vector $\lambda_\ell$, thus requiring more iterations of Grover's algorithm at each stage.
\end{itemize}


When comparing the partial oracle algorithm with the traditional Grover algorithm, some care is needed. Because the optimum case of a partial oracle search bears a resemblance to the case of a binary search on an ordered search space, it is tempting to conclude that any performance comparison is invalid (as Grover's search algorithm presumes the search space is \textit{unordered}). But there are significant differences between partial oracle search and binary search: in a partial oracle search, the ordering in the relationship between the flag bits and the index bits is not given at the outset, but discovered as the search progresses; and the degree of ordering that is discovered falls within a wide range, producing anything from an exponential speed-up to the usual $\sqrt{N}$ oracle queries of Grover's algorithm.

It seems that the partial oracle algorithm delivers search acceleration by discovering correlations between the flag bits and the search index, which represent a kind of order over the search space. In some cases, the discovered order could be simple (for example, if the search space is already ordered), and in other cases, more complex (for example, involving correlations between multiple bits).


In this work, the partial oracle algorithm has been validated for the simplest kind of search problem. Further work is needed to confirm that search acceleration can also be achieved for more complex problems---for example, requiring modelled states that incorporate correlations between two or more bits.

\bibliography{grover}

\end{document}